\documentclass[iop]{emulateapj}
\usepackage{graphicx,natbib,apjfonts}
\usepackage{epstopdf}
\usepackage{arydshln}
\usepackage{amsmath}
\usepackage{gensymb}

\begin{document}

\date{\today}
\title{3d simulations of realistic power halos in magneto-hydrostatic sunspot atmospheres: Linking theory and observation}

\author{Carlos Rijs\altaffilmark{1,2},S.P. Rajaguru\altaffilmark{3},Damien Przybylski\altaffilmark{1,2},Hamed Moradi\altaffilmark{1,2,4},Paul S. Cally\altaffilmark{1,2},Sergiy Shelyag\altaffilmark{1,2}}
\altaffiltext{1}{School of Mathematics, Monash University, Clayton, Victoria 3800, Australia}
\altaffiltext{2}{Monash Centre for Astrophysics (MoCA), Monash University, Clayton, Victoria 3800, Australia}
\altaffiltext{3}{Indian Institute of Astrophysics, Koramangala II Block, Bangalore 560034, India}
\altaffiltext{4}{Trinity College, The University of Melbourne, Royal Parade, Parkville, Victoria 3052, Australia}

\email{carlos.rijs@monash.edu}

\begin{abstract}
The well-observed acoustic halo is an enhancement in time-averaged Doppler velocity and intensity power with respect to quiet-sun values which is prominent for weak and highly inclined field around the penumbra of sunspots and active regions. We perform 3D linear wave modelling with realistic distributed acoustic sources in a MHS sunspot atmosphere and compare the resultant simulation enhancements with multi-height SDO observations of the phenomenon. We find that simulated halos are in good qualitative agreement with observations. We also provide further proof that the underlying process responsible for the halo is the refraction and return of fast magnetic waves which have undergone mode conversion at the critical $a=c$ atmospheric layer. In addition, we also find strong evidence that fast-Alfv\'en mode conversion plays a significant role in the structure of the halo, taking energy away from photospheric and chromospheric heights in the form of field-aligned Alfv\'en waves. This conversion process may explain the observed "dual-ring" halo structure at higher ($> 8 $ mHz) frequencies.
\end{abstract}


\maketitle

\section{Introduction}
A complete picture of the interaction between wave motions and magnetic field in the solar photosphere and chromosphere is not yet available to solar phycisists. \\
Significant uncertainties still exist in the computation of helioseismological inversions in active regions for instance, especially given that the atmosphere above photospheric levels undoubtedly plays a role in muddying the seismic observables at the surface \citep{cally2013}.\\
The theory of mode conversion provides a framework as to how active regions act as a gateway between the subsurface and the overlying atmosphere and modify the properties of otherwise normal acoustic $p$ - modes. \\
The first and most important property of active regions to be explained in a mode conversion context was the well known absorption of $p$ - modes \citep{braun1987}. Upon initial suggestion by \cite{spruit1992}, it was eventually determined that both conversion to the field-aligned slow mode (which travels downwards into the interior) and to the upwards travelling acoustic mode (for non-trapped waves) were the responsible mechanisms \citep{cally1997,cally2003}.\\
The acoustic halo was first noted in Dopplergrams alongside the aforementioned $p$ - mode absorption as a peculiar enhancement in 6 mHz power (with respect to average quiet sun values) which extended several Mm radially outwards from the umbra \citep{braun1992,brown1992,toner1993}. \\
Later, in studies utilising the \emph{Michelson Doppler Imager} (MDI) \citep{scherrer1995} onboard the \emph{Solar and Heliospheric Observatory} (SOHO), it was noted that the enhancement was not present in measurements of the continuum intensity \citep{hindman1998,jain2002}. \\
This suggests that either there is a process at work affecting observed power somewhere in the height range between the intensity continuum height and the Doppler velocity observation height, or that the mechanism causing the enhancement is not a process that is detectable in measurements of intensities. \\
It turns out that the former case is much more likely, as intensity halos taken from spectral lines at greater heights have since been observed and studied in detail \citep{moretti2007,rajaguru2012}.\\
Also using MDI, \citet{schunker2011} examined 7 days of observations of the active region AR 9787 and showed that halos are manifested for relatively horizontally aligned, weak-to-moderate magnetic fields ($150$ G $< |\mathbf{B}| < 350$G). This study also noted the interesting property that the power spectrum ridges of the enhancement region were shifted towards a larger wavenumber ($k$) for a given frequency ($\nu$) (compared to the ridges from an area of the quiet sun) and that this effect is more pronounced for larger $k$, which sugggests that shallower waves are being more strongly affected in the enhancement region. \\
The most comprehensive observational halo study to date, by \citet{rajaguru2012} utilised the \emph{Helioseismic and Magnetic Imager} (HMI) and \emph{Atmospheric Imaging Assembly} (AIA) instruments onboard the \emph{Solar Dynamics Observatory}. \\
The authors conducted a multi height analysis of several active regions, measuring the time-averaged power from intensities and velocities corresponding to 6 different heights. From the intensity continuum at $z=0$ (the base of the photosphere, where the optical depth is  unity) to Doppler velocities of the Fe I 6173.34 {\AA} line at around $z=140$ km to intensities measured from the AIA 1600 {\AA} and 1700 {\AA} chromospheric spectral lines, halo properties were compared and analysed in detail. The findings can be summarised as follows:
\begin{enumerate}

\item The halo is present for non-trapped frequencies, beginning at 5.5 - 6 mHz (as observed by all references above) and is present up to at least 9-10 mHz. The 6 mHz halo is the strongest in measurements of the Fe I 6173.34 {\AA} Doppler velocity at $z=140$ km.
\item The halo magnitude is a clear function of height. There is no enhancement in the time-averaged intensity continuum ($z=0$) power or in the derived line-wing Doppler velocity ($z=20$ km). For weak-field regions at these heights, there is also a uniform wave power above the acoustic cutoff, which is to be expected. However, at $z=140$ km (the aforementioned HMI Doppler velocity line) the situation is markedly different, and the halo comes into effect. \\
\item The halo is clearly present in observations of the chromosphere, as measured by AIA. The time-averaged power of the 1600 {\AA} and 1700 {\AA} wavelength channels (corresponding approximately to $z=430$ and $360$ km respectively) shows a halo in the 7-10 mHz range, that spreads radially with height, agreeing with the suggestions of \citet{finsterle2004}.
\item The spatial extent and structure of the halo changes above about 8 mHz. This higher frequency halo is seen in power maps to be thinner and more confined spatially than the more diffuse structure seen at 6 mHz. Radially outwards from this higher $\nu$ field is a region of slightly reduced power, which in turn is surrounded by a diffuse, weak halo region, extending radially many Mm into quiet regions \citep{rajaguru2012}.\\

\end{enumerate}

In this study we are interested in providing a consistent theoretical explanation for the acoustic halo. There are of course a variety of existing theories as to the mechanism behind the phenomenon. \\
By conducting radiative simulations for instance, \citet{jacoutot2008} determined that strong magnetic fields can alter the scale size of granulation cells, which in turn can modify the local excitation frequency of resultant photospheric waves. They found that the stronger field also increases the amplitude of non-trapped waves at frequencies consistent with halos. \\
\citet{kuridze2008} show semi-analytically that waves with $m>1$ (where $m$ is the azimuthal wave number) can become trapped under field free canopy regions, resulting in an enhancement of higher frequency wave power.\\
\citet{hanasoge2009} suggests that the halo is a consequence of the equilibrium state of the solar surface, and that the local oscillation can be shifted to a lower mode mass \citep{bogdan1996} due to scattering from the magnetic flux tube. \\
We will discuss why these theories do not appear viable in light of our simulation results in the discussion at the end of this paper.\\

\subsection{Mode conversion}
In \citet{rijs2015}, we performed 3D simulations to determine whether there was promise in the suggestion of \citet{khomenko2009} that it is in fact the \emph{overlying} atmosphere that is directly responsible for the halo. Specifically that the addition of energy from high frequency non-trapped waves which have travelled above the Alfv\'en-acoustic equipartition ($a=c$) layer and undergone mode conversion and refraction are responsible. In this case, the process of mode conversion describes the intrinsic physics. \\
At greater depths below the solar photosphere, the plasma $\beta$ (where $\beta = P_{g}/P_{m}$, with $P_{g}$ and $P_{m}$ being the gas and magnetic pressures respectively) increases. Several Mm below the surface the plasma is dominated by hydrodynamic physics and waves are governed by the standard gas sound speed ($c$). \\
Conversely (assuming one is in the proximity of an active region of some sort), well above the surface the gas density ($\rho$) drops, the $\beta$ becomes small and waves are governed more strongly by the Alfv\'en speed ($a$), where $a \propto |\mathbf{B}|/\sqrt{\rho}$, and |$\mathbf{B}$| is the local magnetic field strength.\\
There is therefore a layer of the atmosphere (roughly where $\beta=1$) where $a$ and $c$ equate - the so called $a=c$ layer. At this height, the phase speeds of the magnetoacoustic fast and slow waves are equal, allowing the two modes to interact. Energy can be channeled from the fast to the slow branch or vice versa. \\
The fast wave is largely acoustic in nature when $a<c$ and magnetic when $a>c$, and it is this fast magnetoacoustic wave that will refract and then reflect at the fast wave turning height (where $\omega^{2}=a^{2}k_{h}^{2}$, with $\omega=2 \pi \nu$ and $k_{h}$ being the horizontal component of the wavenumber, $k$), returning downwards from above the $a=c$ layer. \\
Energy is preferentially converted from the fast-acoustic mode to the fast-magnetic mode if there is a large attack angle between the wavevector of the incident wave and the orientation of the magnetic field. If the attack angle is small then energy will be primarily channeled into the field aligned slow mode. This perhaps explains why halos are observed amongst horizontal field; The line of sight component of the Doppler velocity is largely vertical (when observing at disk center) and provides a large attack angle with the horizontal field.\\
The theory can also explain the spreading of the halo that is observed with height \citep{rajaguru2012}, given that the $a=c$ layer is located at greater radial distances from the umbra as a function of height.\\
Waves with frequencies below the acoustic cut-off are generally unable to reach the $a=c$ height, as they have reflected inwards, which is presumably why halos are only observed at non-trapped frequencies. The fast magneto-acoustic wave provides the excess energy in observable regions, which would otherwise not be present in the quiet sun (see \citealt{cally2006,schunker2006} for further details on mode conversion or \citealt{cally2007} for a succinct review of the theory).\\
\citet{khomenko2009} have performed simulations with both monochromatic and gaussian wave sources in a magneto-hydrostatic (MHS) sunspot atmosphere and show a clear correlation between the power halo and a suspicious increase in RMS velocities for non-trapped waves resulting from the interference pattern generated by downwards travelling fast waves. \\
In \cite{rijs2015} we extended this work in 3D. By performing forward modelling simulations with a spatially localised gaussian (in space, time and frequency) wave pulse, the halo structure resulting from the vertical component of velocity ($v_{z}$) was analysed as a function of radius, height and frequency. A clear correlation between the position of the $a=c$ layer and the halo was shown and the dependancy of the halo on the overlying atmosphere was exhibited.\\
In this work we perform simulations in similar MHS sunspot atmospheres to those of \citet{rijs2015}. However we now use a realistic distributed wave source, modelled as a slab of point sources at some depth below the photosphere. The sources are tuned to mimic the observed photospheric power spectrum, peaking at the 5 minute oscillation period ($\nu=3.3$ mHz) and exhibiting solar-like amplitudes. In this way we are able to compare the halos present in our simulations with observations in a more rigorous manner. \\
For the observational comparisons we use a subset of the data corresponding to a single active region from \citet{rajaguru2012} which provides a multi-height velocity and intensity halo data set with which to compare our simulations. 


\section{The Simulation}

In this section we present an overview of our simulations, including the details of the sunspot atmosphere used, a summary of the distributed wave source and details regarding the calculation of synthetic instensities, phase shifts and velocities.

\subsection{The MHS atmosphere}
A detailed description of the sunspot model we are using can be found in \citet{przybylski2015}, where the model of \citet{khomenko2008} is optimised in order to increase spectropolarimetric accuracy and produce more realistic line formation regions.\\
In short, the MHS configuration combines the self-similar sub-photospheric model of \cite{low1980} with the potential configuration of \cite{pizzo1986}. Convective stability is enforced by the method of \cite{parchevsky2007}. \\
The model makes use of the Model S for the distribution of quiet subphotospheric thermodynamic variables \citep{dalsgaard1996} and the Avrett umbra for the non-quiet variables \citep{avrett1981}. The VALIIIC chromosphere \citep{vernazza1981} is smoothly joined onto these distributions to complete the full model, yielding a sunspot-like magnetic field configuration embedded into the atmosphere.\\
The sunspots we use in this instance are similar to those used in \cite{rijs2015} and \cite{moradi2015}, except for some parameters, such as the peak field strength, the inclination at the surface, and the simulation box size, which have been modified. \\
The sunspot model not only provides the freedom to choose the peak field strength at the surface of the photosphere but also the depth of the Wilson depression (the height at which the atmosphere becomes optically thin is depressed in high field regions such as the umbra). As such we make use of two model atmospheres in this study, one with a peak surface field strength of $|\mathbf{B}|=1.4$ kG and another with $|\mathbf{B}|=2.7$ kG. The atmospheres have Wilson depression depths of 250 and 400 km respectively, which are reasonably realistic values. \\
The surface of the atmosphere is defined as the photospheric height at which $\log(\tau)=0$ (where $\tau$ is the optical depth scale, as calculated from the known thermodynamic values at every point in the box) and follows the Wilson depression. The surface corresponds to the height of formation of the 5000 {\AA} intensity continuum ($z=0$) in this atmosphere.\\

\subsection{Forward modelling}
\begin{figure*}
\centering
\includegraphics[width=\textwidth]{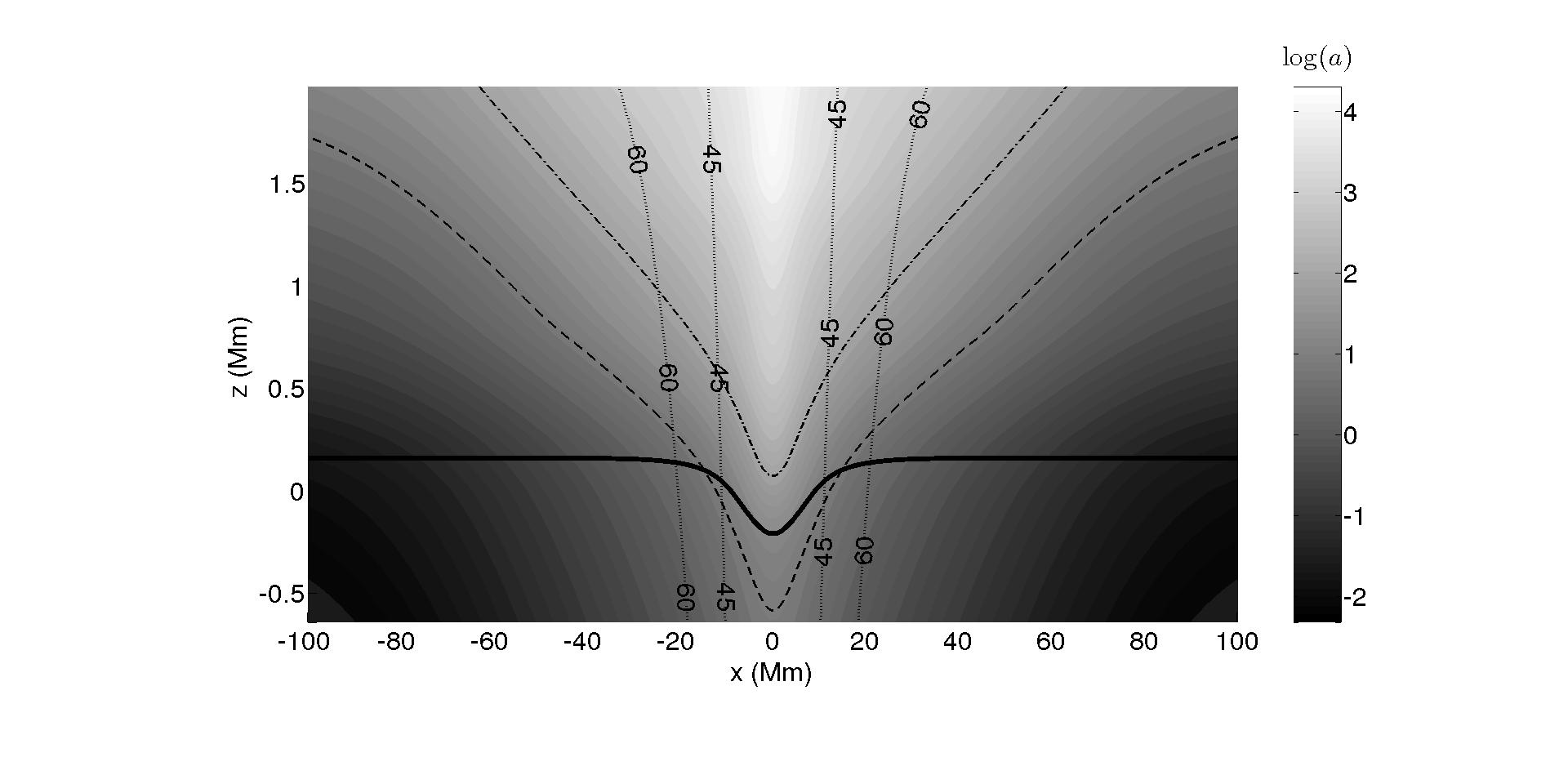}
\caption{A cut through the sunspot center. Field inclination contours are shown for typical umbral/penumbral and penumbral/quiet sun values of 45 and 60 degrees from the vertical respectively. The surface or photosphere layer, where $\log(\tau)=0$, is shown by the solid black curve. The dashed curve is the $a=c$ equipartition layer for this atmosphere and the dash-dotted curve is the $a = 90$ km/s layer, where the Alfv\'en limiter is in effect. The background contour is $\log(a)$ in km/s as it would appear without any Alfv\'en limiter in application. In our simulations $a$ is constant above the $a=90$ km/s curve. Note that the aspect ratio of the figure is highly stretched, with the horizontal dimension spanning 200 Mm and the vertical spanning only around 2 Mm.}\label{sim1}
\end{figure*}

As in our previous work, we use the SPARC code for forward modelling \citep{hanasogephd2007, hanasoge2007}. The code has been used several times for wave-sunspot interaction studies \citep{moradi2013,moradi2014,moradi2015}.\\
The code solves the ideal linearised MHD equations in cartesian geometry. \\
As input, we define a background atmosphere and instigate wave propagation for the desired simulation length. The background atmosphere can be any magnetic plasma such as the sunspot atmospheres mentioned above or any quiet-sun atmosphere, provided it is convectively stable. \\
The output is the perturbations to the background states of the pressure ($p$), $\rho$, magnetic field ($\mathbf{B}=[B_{x},B_{y},B_{z}$]) and velocity ($\mathbf{v}=[v_{x},v_{y},v_{z}]$).\\ 
The computational domain in both cases is square in the horizontal with 256 points in each of the $x$ and $y$ directions (where $L_{x}=L_{y}=200$ Mm) yielding a horizontal spatial resolution of $\delta x=0.78125$ Mm. 
There are 220 grid points in the vertical direction, with spacings scaled by the local background sound speed. This results in vertical grid spacings of around 20 km near the surface and 100 km at depths of several Mm. The box extends from a depth of 10 Mm below the surface to 2.5 Mm above it in this manner.\\
Side boundary conditions in our simulations are periodic, and there are both damping sponges and perfectly matched layers (PML) in effect along the top and bottom boundaries of the box. The top 20 and the bottom 8 grid points are taken up by these sponges and the PML, resulting in a maximum useable box height of 2 Mm (above the surface).\\
In order to overcome the numerical challenges of explicit forward modelling in an atmosphere where the governing wave speed scale (the Alfv\'en speed, $a$) increases rapidly above the surface, we use the Alfv\'en speed limiter described by \citet{rempel2009}, which was also used in \cite{rijs2015}. This allows us to sidestep the requirement of using a prohibitively small simulation time step, imposed by the Courant-Friedrichs-Lewy (CFL) condition. Work has been done to ascertain whether the use of an Alfv\'en speed limiter has a detrimental effect on helioseismic travel times \citep{moradi2014}, with the conclusion being that one must be certain that the artificial capping is occuring well above heights where any relevant physics is occuring (such as the fast wave reflection height or the $a=c$ layer). \\
We have set our limiter at a value of $a_{lim}=90$ km/s, yielding a simulation time step of around 0.2 seconds. \\
Figure \ref{sim1} shows a cut through the centre of our 2.7 kG sunspot atmosphere (along the plane at $y=0$). Overlaid are the $a=c$ and $a=90$ km/s contours, as well as the photospheric surface, with a Wilson depression of around 400 km.\\
The vertical inclination contours show the rather rapid drop-off in field inclination, with the field reaching 30 degrees from the horizontal some 20 Mm from the umbra (at the surface).\\
To reiterate, the mode conversion effects occur around the $a=c$ layer, and so it is important that fast waves are given space to refract back downwards as they naturally would before the limiter at $a=90$ km/s takes effect. We have taken care to ensure that this is the case and that the modification of the atmosphere will not affect these returning fast waves. In this regard, simulations have been run with Alfv\'en limiter values up to 200 km/s, with no change to the halo properties observed.\\
\begin{figure*}
\centering
\includegraphics[width=\textwidth]{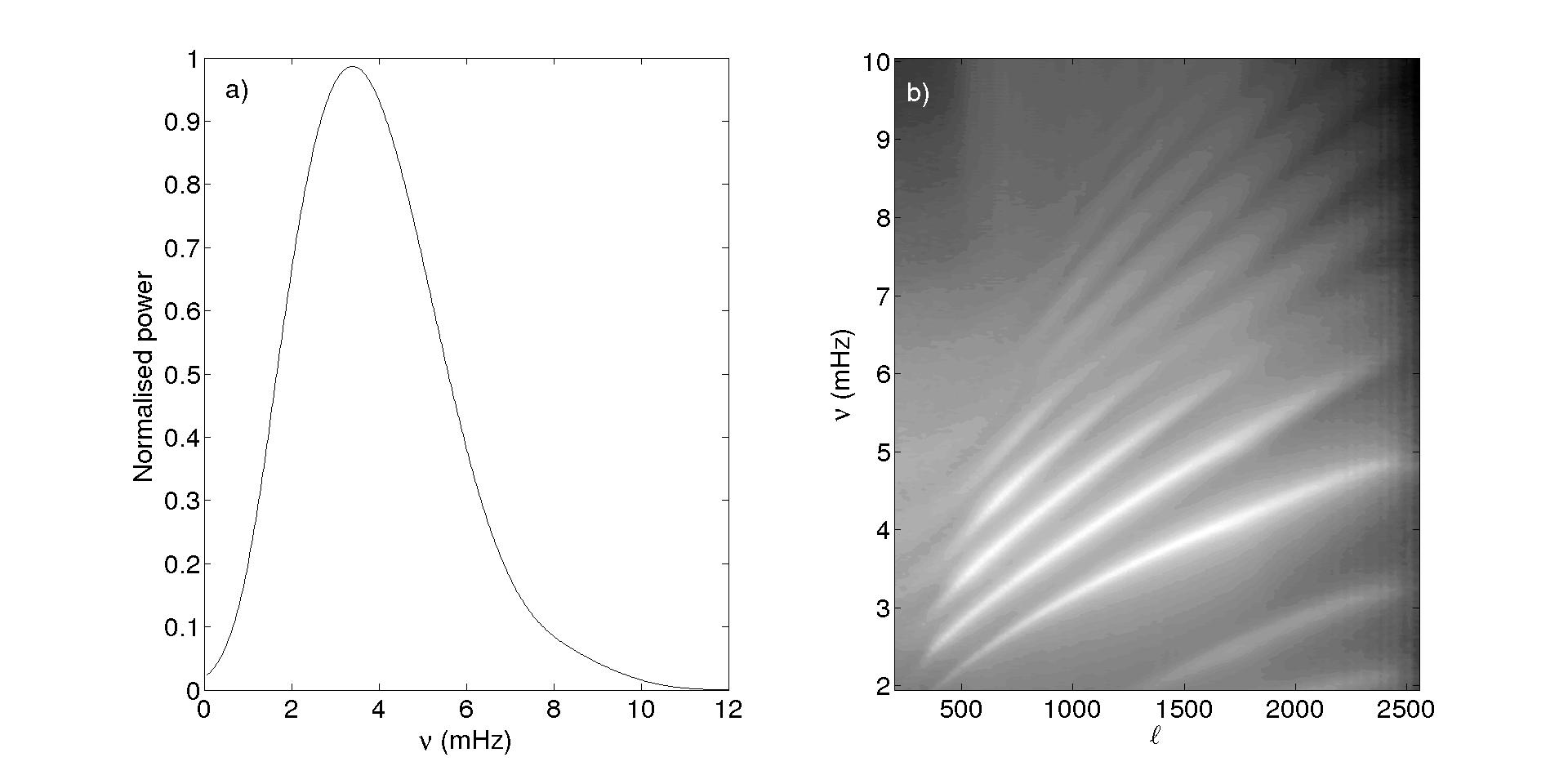}
\caption{Panel a): The power spectrum of the wave source function used, tuned to provide a solar-like peak. b:) Arbitrarily normalised power ridges in $\ell$-$\nu$ space for 6 hours of simulation time, calculated at the surface ($z=0$) from $v_{z}$.}\label{sim2}
\end{figure*}
Regarding our wave source function, we are attempting to model the uncorrelated stochastic wave field seen on the solar photosphere. This wave field is, in reality, generated by subsurface convective cells. We choose a depth of 150 km below the surface and initiate a source function, $S$, in the manner of \cite{hanasoge2007}, i.e.
\begin{equation}
S(x,y,z,t)=\hat{S}(x,y,t) f(z)
\end{equation}
where the horizontal function $\hat{S}(x,y,t)$ is a plane of spatial delta functions which are excited at a cadence of 30 seconds, and the function $f(z)$ is a gaussian function in depth with FWHM of approximately 100 km centered at 150 km below the surface.\\
The source power spectrum has been tuned such that it more-or-less fits the spectrum of power observed on the surface of the quiet sun. Panel a) of Figure \ref{sim2} displays this spectrum, with a peak in power at around 3.3 mHz, and non zero power present until above 10 mHz. Panel b) shows the power ridges in $\ell$-$\nu$ space calculated from 6 hours of $v_{z}$ output at the surface.\\
In taking into account the fact that strong umbral fields inhibit subsurface convection and wave propagation, we do not excite waves in the umbra of the sunspot itself, smoothly suppressing the source amplitude as the magnetic field strength increases.\\ 
Wave propagation is initiated and run for 6 hours of solar time in total using both the 1.4 kG and 2.7 kG sunspot atmospheres (as separate simulations).\\
We analyse the power manifested in synthetic intensities corresponding to the 5000 {\AA} continuum intensity, the AIA 1700 {\AA} and 1600 {\AA} intensity bands as well as both the vertical and horizontal components of the velocity perturbation ($v_{z}$ and $v_{h}$ respectively), which correspond observationally to the line-of-sight components of velocity when observing at disk centre ($v_{z}$) and at the limb ($v_{h}$).\\
In reality, the HMI Doppler camera \citep{scherrer2012} measures velocities from the Fe I 6173.34 {\AA} line, which has its peak of formation at a height of around 140 km \citep{fleck2011,rajaguru2012}, while the AIA \citep{lemen2012} 1700 {\AA} and 1600 {\AA} wavelength intensity channels are formed at approximate heights of 360 km and 430 km respectively \citep{fossum2005,rajaguru2012}. \\
Thus, in comparing the structure of power enhancements present in our simulations with the observed power behaviour from \cite{rajaguru2012} we extract simulation velocity signals from a height of $z=140$ km. We then calculate the synthetic intensities corresponding to the two above AIA channels as well as the 5000 {\AA} continuum intensity for our 6 hour wave propagation simulations. The approximate 1600 {\AA} and 1700 {\AA} intensities are calculated by interpolating the ATLAS9 continuum and line opacity tables \citep{kurucz1993} using the plasma parameters from the simulation and integrating them together with the corresponding LTE source function along the lines-of-sight for each column in the sunspot models. The routine used for the intensity calculation is similar to that of \cite{jess2012}. The filter bandwidths are set to 10 {\AA} for both simulated AIA channels. No line-of-sight velocity or magnetic field information is used in this radiation intensity calculation.

\section{Comparisons with observations - vertical velocities and intensities}
\begin{figure*}
   \centering
    \includegraphics[width=1.0\textwidth,trim=0cm 0cm 0cm 1cm]{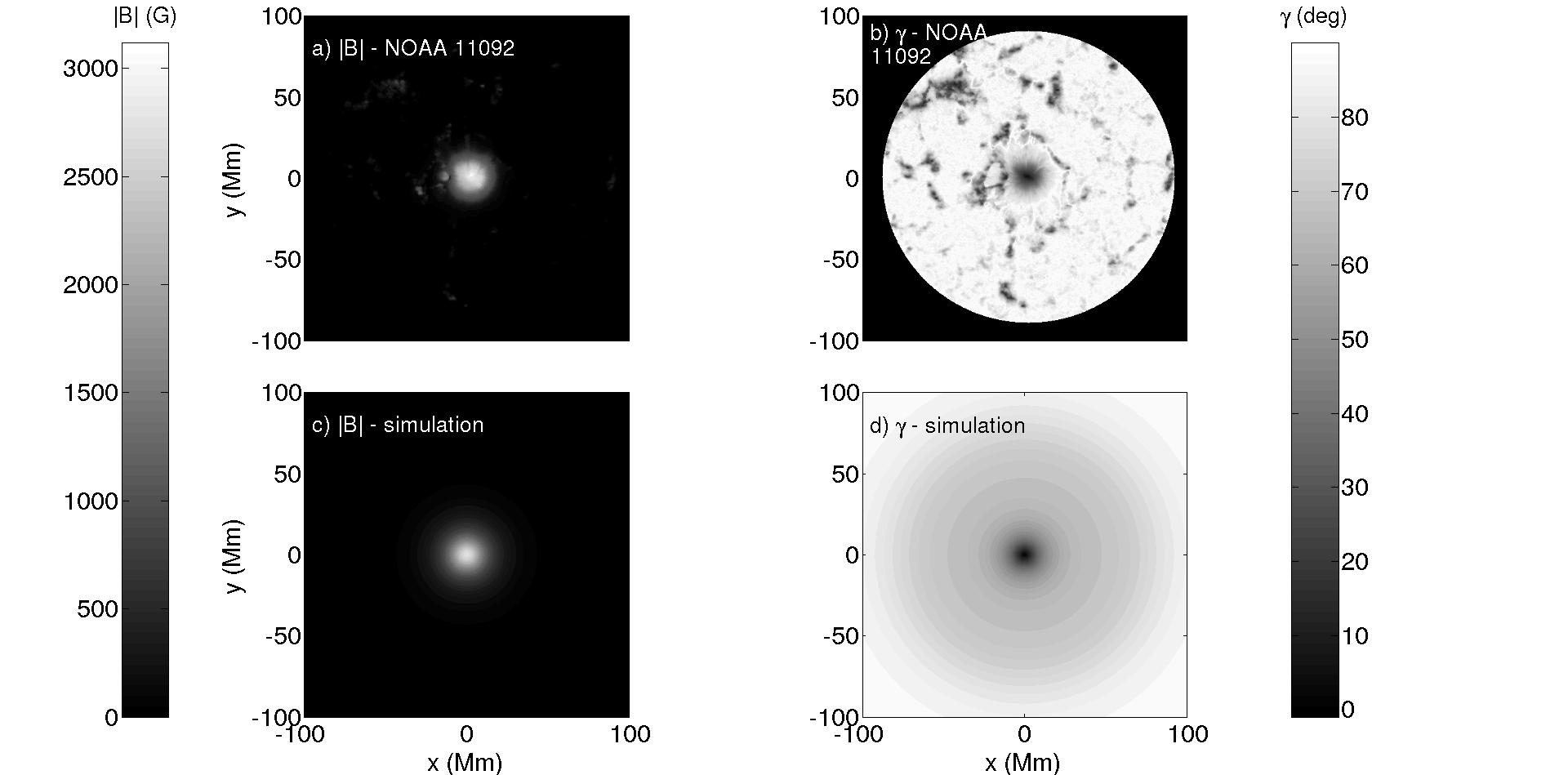}
    \caption{Panels a) and b) show |$\mathbf{B}$| and the unsigned field inclination from vertical ($\gamma$) respectively for NOAA 11092. Panels c) and d) are the counterpart plots for the 2.7 kG simulated sunspot atmosphere.}\label{sim-obso}
\end{figure*}

\begin{figure*}
   \centering
    \includegraphics[width=1.0\textwidth,trim=0cm 0cm 0cm 1cm]{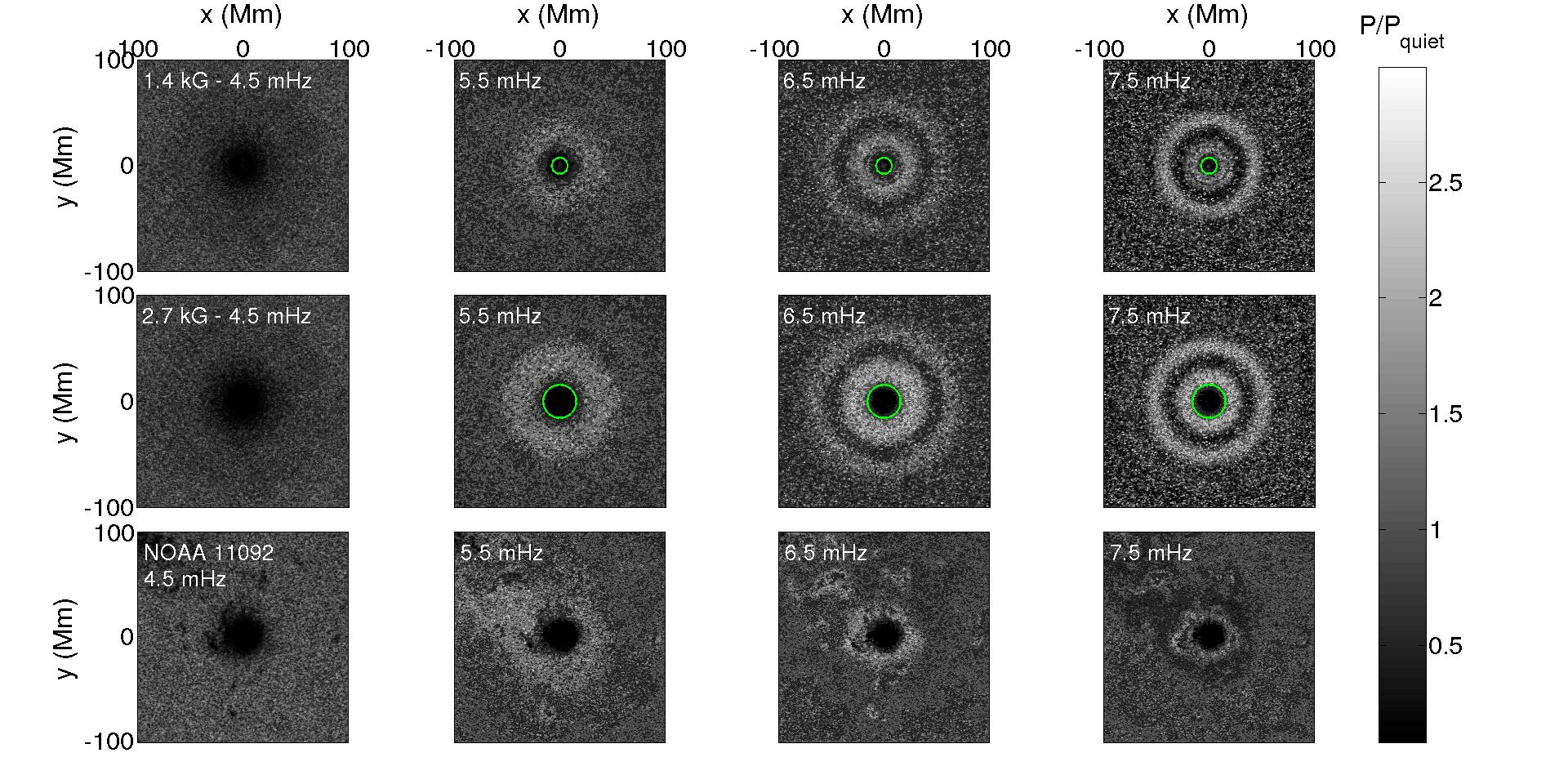}
    \caption{Top row - 6 hr time-averaged $v_{z}$ power maps at the height of formation of the Fe 6173.34 {\AA} line ($z=140$ km) for 4 illustrative frequency ranges for the weak sunspot case (1.4 kG). Middle row - The same power maps for the stronger field case (2.7 kG). Bottom row - 14 hr time averaged observational Doppler velocity power maps of the active region NOAA 11092 for the same frequency
ranges. The green contour in rows 1 and 2 is the $a=c$ contour at $z=140$ km. }\label{obso1}
\end{figure*}

\begin{figure*}
   \centering
    \includegraphics[width=1.0\textwidth,trim=0cm 1cm 0cm 1cm]{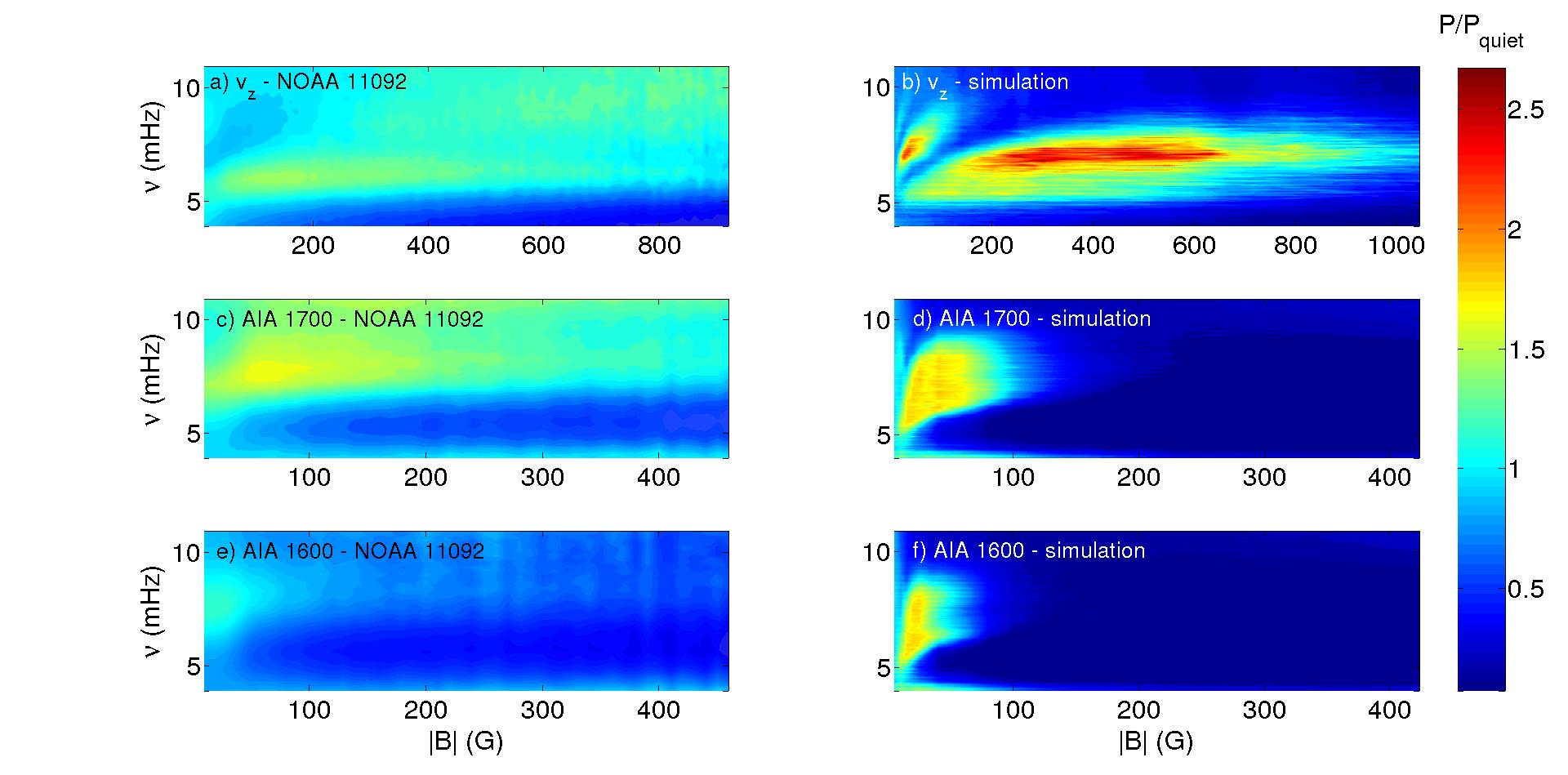}
    \caption{Panel a): Unfiltered Doppler velocity power as a function of $|B|$ and $\nu$. b): 2.7 kG simulation unfiltered $v_{z}$ power. c) \& d): observed and 2.7 kG synthetic unfiltered AIA 1700 power respectively.  e) \& f): observed and 2.7 kG synthetic unfiltered AIA 1600 power respectively.}\label{obso2}
\end{figure*}

\begin{figure*}
   \centering
    \includegraphics[width=1.0\textwidth,trim=0cm 1cm 0cm 1cm]{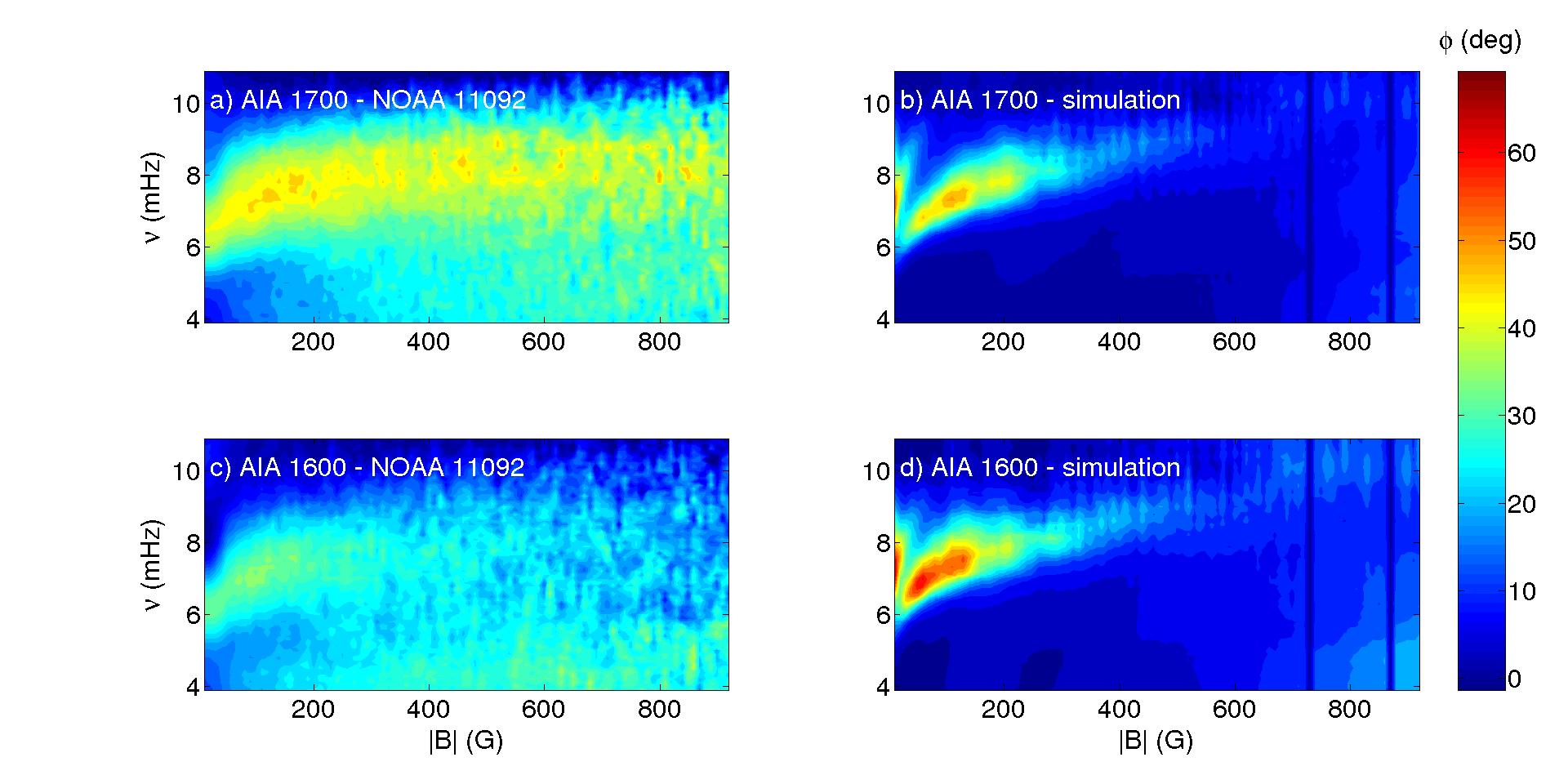}
    \caption{Phase shifts at the heights of formation of the AIA 1700 {\AA} and 1600 {\AA} lines of all waves with respect to those at the surface. Panels a) and c) correspond to observations and b) and d) to the 2.7 kG simulation.}\label{obso3}
\end{figure*}

This section details the comparisons between the power structures present in our 6 hour simulations and those observed in the active region NOAA 11092.\\
As shown in observations, the acoustic halo is a phenomenon especially sensitive to |$\mathbf{B}$| and to the local field inclination. We firstly demonstrate here some of the similarities and differences in these properties exhibited by the artificial sunspots and the real active region.\\
Figure \ref{sim-obso} compares the topology of |$\mathbf{B}$| and the unsigned field inclination from the vertical ($\gamma$) at the surface our 2.7 kG sunspot atmosphere and NOAA 11092.\\
For the observations of NOAA 11092, |$\mathbf{B}$| is calculated from the disambiguated vector maps with components $B_{x}$, $B_{y}$ and $B_{z}$, where |$\mathbf{B}$|$=\sqrt{B_{x}^{2}+B_{y}^{2}+B_{z}^{2}}$. $\gamma$ in degrees is then simply defined as $\gamma=90-(180/\pi)|\arctan(B_{z}/B_{h})|$ where $B_{h}=\sqrt{B_{x}^{2}+B_{y}^{2}}$.\\
As can be seen in the figure, the field strength of NOAA 11092 drops off in a similar manner to the artificial sunspot, however the small scale features present in the real active region introduce many variations in both the field and its inclination which are not modelled in our simulations. The behaviour of $\gamma$ around NOAA 11092 with radius for example is not the smooth monotonically increasing function yielded by the 2.7 kG sunspot model. We can therefore expect some differences between observed and simulated halo structure will result.\\
Firstly, we compare the acoustic power for $v_{z}$ - from both the weak (1.4 kG) and strong (2.7 kG) sunspot atmospheres - with the 14 hour time averaged Doppler velocity power from NOAA 11092.\\
Power maps are shown in Figure \ref{obso1} for a range of frequencies of interest. The power at every point has been divided by the average power of a quiet corner of the simulation domain, in order to represent an enhancement over quiet values.
In both simulations, the enhancement comes into effect at around 5 mHz, when waves are in the non-trapped regime, just as in the observations.\\
The differences between the two sunspot simulations (rows 1 and 2) are immediately evident, with the 2.7 kG sunspot exhibiting a larger umbra. A consequence of having a stronger magnetic field strength is also that the $a=c$ height will be lower in the atmosphere, resulting in a spreading of this contour for a particular observation height. It is clear that the halo appears correlated with the $a=c$ contour in both cases.\\ 
An intriguing feature of the simulated halos is the clear dual-ring structure present for higher frequencies. The inner ring appears to conform qualitatively well at a glance with the observational halo. However the rings appear to be interrupted by a region of mild power deficit (with respect to the quiet sun).\\
Although not immediately visible in the power maps in the bottom row of Figure \ref{obso1}, observed halos do exhibit a similar structural change when observed at increasingly high frequencies. This feature can clearly be seen in power maps of observed Doppler velocity in \citet{rajaguru2012} and \citet{hanson2015} at 8 and 9 mHz respectively.\\
In section 5 of this paper we discuss how fast-Alfv\'en conversion likely leads to this dual-ring structure.\\
Comparing power maps in this way is of only so much use. To more rigorously compare the structure of observed and simulated power halos we plot unfiltered power enhancements as functions of |$\mathbf{B}$| and $\nu$ (i.e. no frequency filter is applied during the fourier transform.) In this way we may fully examine the spectral structure of the halo (Figure \ref{obso2}).
\\ 
Also present in Figure \ref{obso2} is the power calculated from the AIA 1700 {\AA} and AIA 1600 {\AA} intensity bands, which we have synthetically calculated in our simulations in order to compare to observations.\\
The left hand panels show the NOAA11092 power structure for the Doppler velocity and intensities, whereas the right hand panels correspond to simulation output for the 2.7 kG sunspot atmosphere. To be clear, panel b) of Figure \ref{obso2} corresponds directly to the power maps in the middle row of Figure \ref{obso1}, it is simply unfiltered in frequency space and so is inclusive of the entire spectral structure. The power at every point has been binned according to the local value of $\mathbf{|B|}$ and then averaged so as to reveal not only the spectral structure of the halo but also how it behaves with respect to field strength. \\
The first thing to notice in Figure \ref{obso2} is that the simulated $v_{z}$ power structure (panel b) matches up reasonably well with the observed Doppler power (panel a). The halo has formed over relatively weak field ( 50 G < $|\mathbf{B}|$ < 700 G) as expected. \\
In the simulation, $|\mathbf{B}|$ decreases (and $\gamma$ increases) smoothly and uniformly as one moves away from the umbra. As such this field strength range corresponds to nearly horizontal inclinations of 55\degree < $\gamma$ < 75\degree. \\
This seems to also agree with all other observational reports of enhancements which place the halo amongst moderate to weak and horizontally inclined field \citep{jain2002,schunker2011,rajaguru2012}.\\
The dual-ring structure can clearly be seen at higher frequencies in panel b), manifesting as the second lobe of enhancement for very weak field. Wedged between the two rings (at around $(|\mathbf{B}|,\nu)=(100,6)$ is the clear region of power reduction.\\
Looking at greater heights in the form of the AIA 1700 {\AA} and 1600 {\AA} intensities (corresponding to heights of 360 km and 430 km above the base of the photosphere respectively) we also see a general agreement in $\nu,|\mathbf{B}|$ space. The spreading of the magnetic canopy at these heights has resulted in the intensity halos forming at much weaker field locations both in the observations and the simulations. \\
The magnitudes of the enhancements in the simulations are consistently larger than the observed values, as evident from this figure. This is a feature that was also noted in \citet{rijs2015} and can most likely be attributed to the fact that our sunspot is symmetric and its magnetic field inclination is a steep, monotonically decreasing function of radial distance. The MHS structure is such that horizontal field is enforced at the side boundaries of the simulation domain and so there is a large expanse of nearly horizontal field. As explained previously, the fast-slow mode conversion mechanism for the generation of the halo relies on a large attack angle between wavevector and field and so, in analysing $v_{z}$ power enhancements, it is reasonable to expect that this horizontal field will be very conducive to the conversion of energy into magnetic fast waves and hence, a prominent halo. \\
Power derived from the 5000 {\AA} intensity continuum (at $z=0$) was also calculated synthetically to compare with the observational intensity continuum power. It is well known that halos do not appear in measurements of intensity continuum power and we also found this to be the case, with no enhancement present. \\
Another intersting result, shown in Figure \ref{obso3}, is the comparison between observed and simulated phase shifts.
A net upward or downward propagation of waves in an atmosphere can be diagnosed by calculating the temporal cross-spectrum of any wave quantity sampled at two different heights. \\
For example, for velocities $ v(z_{1},t)$ and $v(z_{2},t)$ sampled at two different heights $z_{1}$ and $z_{2}$, the phase shift corresponding to a height evolution of the wave is given by the argument or phase of the complex cross-spectrum, 
\begin{equation}
\phi_{1,2}(\nu)=\text{arg}[{\mathbf V}(z_{1},\nu){\mathbf V^{*}}(z_{2},\nu)],
\end{equation}
where ${\mathbf V}$ is the Fourier transform of $v$. In the above convention, a positive phase-shift would mean that the wave is propagating from height $z_{1}$ to $z_{2}$, while the opposite holds for a negative phase-shift.\\
The phase shift contour maps of Figure \ref{obso3} describe the phase shifts of waves at the AIA 1700 {\AA} and 1600 {\AA} intensity formation ($z=360$ km and $430$ km respectively) with respect to those at the height of formation of the intensity continuum. \\
The simulation yields a clean band of positive phase shifts at halo frequencies with respect to those at the surface at weak field regions. The same basic pattern is seen in the observations, however there is some extended phase shift structure at higher field strengths in the AIA 1700 {\AA} power (panel a) which is not replicated in the simulation.\\
The simulation phase shifts are also of a greater magnitude than observations - particularly in the case of the AIA 1600 intensities. \\
These variations in features are not too surprising. Considering Figure \ref{sim-obso} we see that NOAA 11092 exhibits a much more rapid horizontality of field away from the umbra than seen in the MHS model. We show in section 5 how these bands of positive phase shifts at given observation heights may be intrinsically related to the process of fast-Alfv\'en mode conversion. The physics of fast-Alfv\'en mode conversion are strongly tied to the local magnetic field inclination. Therefore the reason that NOAA 11092 exhibits such an extended phase shift structure into higher field regions (and our MHS sunspot does not) may be in part due to the more horizontal field at those radii for the active region.

\section{The theoretical underpinnings of Halos}

\begin{figure*}
   \centering
    \includegraphics[width=1.0\textwidth,trim=0cm 1cm 0cm 1cm]{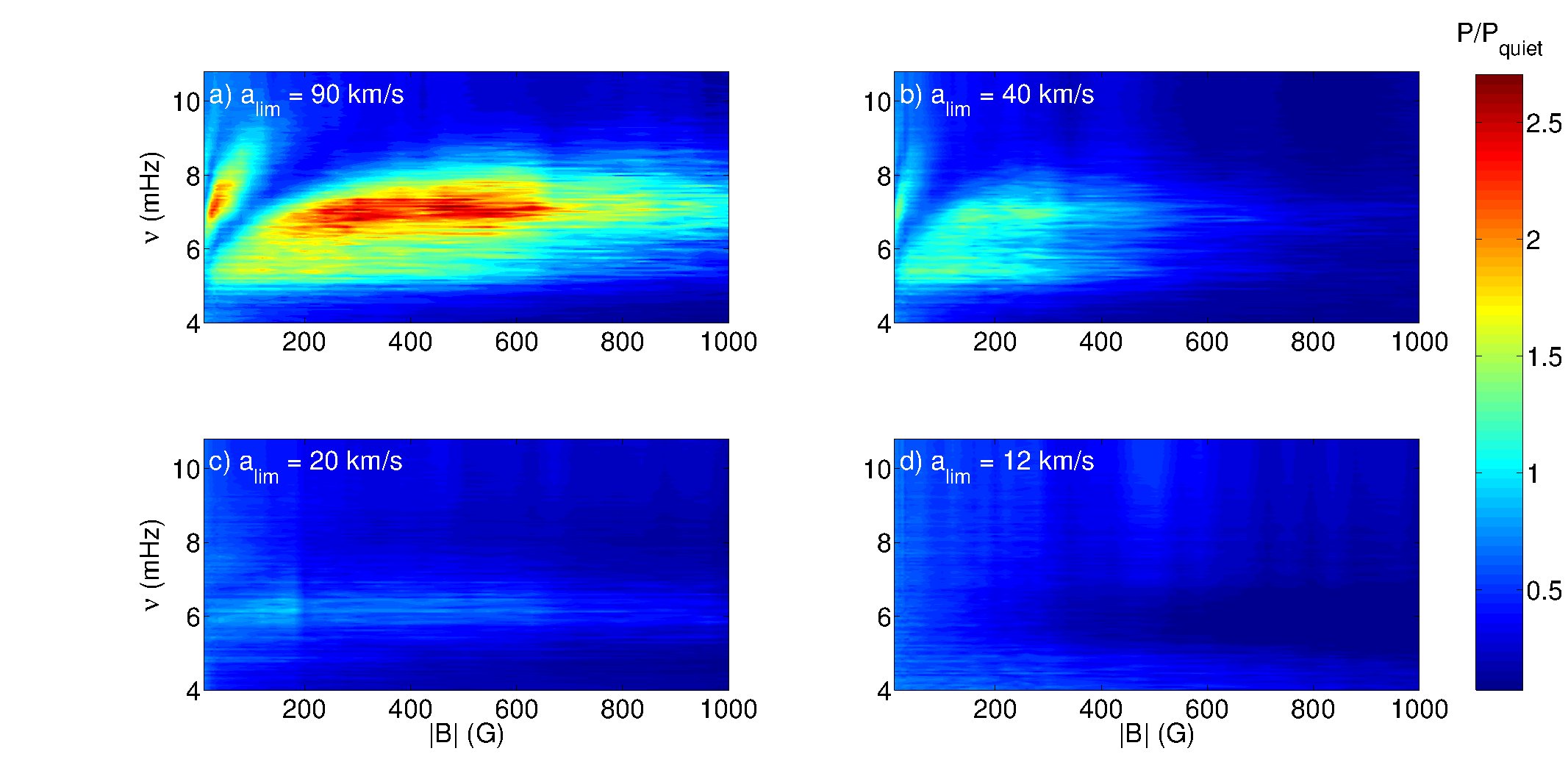}
    \caption{Panel a): Unfiltered $v_{z}$ power halos in the case $a_{lim}=90$ km/s. b), c) and d) show the same quantity from simulations with progressively lower values of $a_{lim}$.}\label{theory1}
\end{figure*}

\begin{figure*}
   \centering
    \includegraphics[width=1.0\textwidth,trim=0cm 1cm 0cm 1cm]{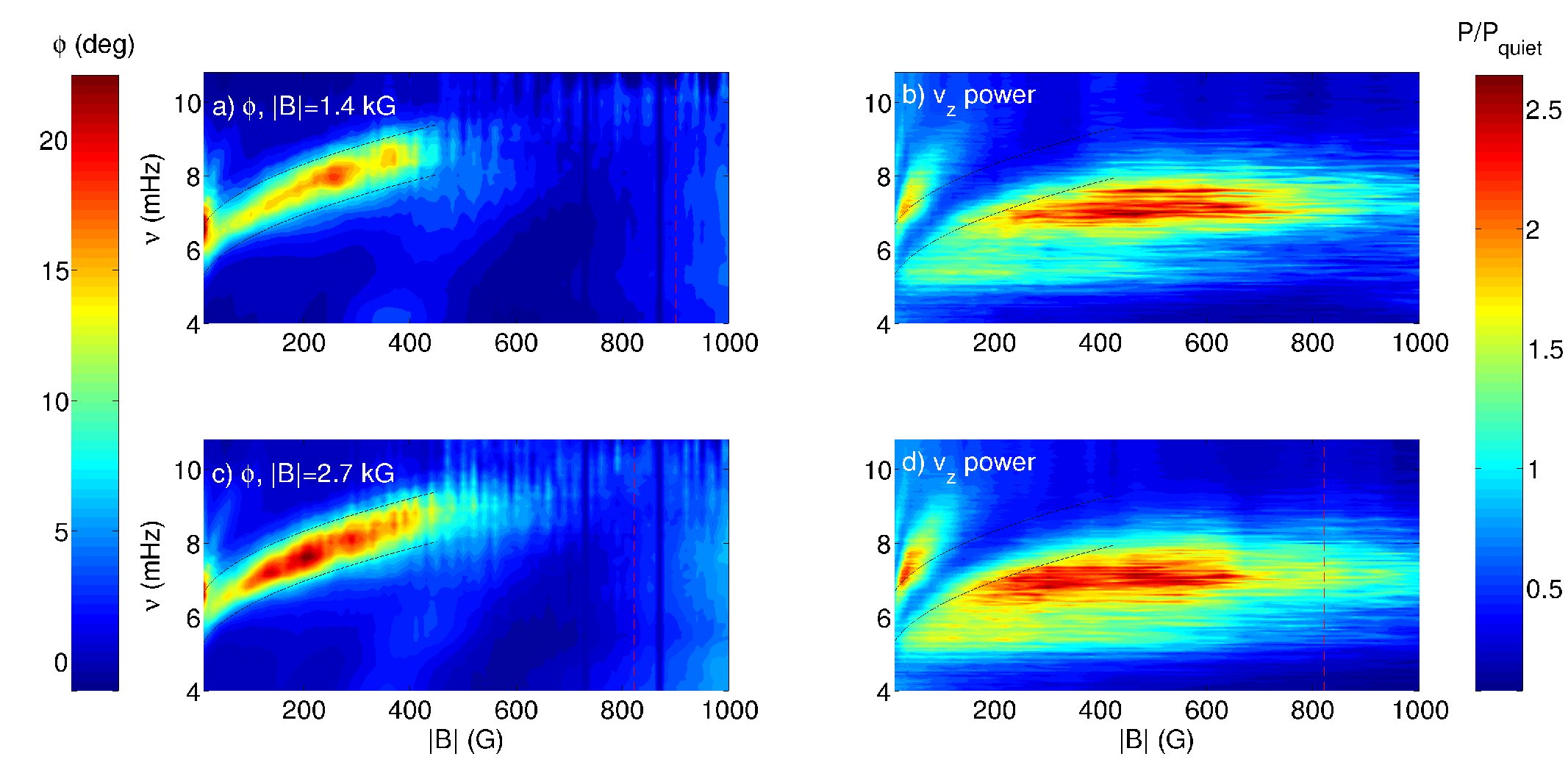}
    \caption{The top row corresponds to the weak field simulation, with peak field strength 1.4 kG, and the bottom row is from the strong field case (2.7 kG peak). On the left are phase shifts calculated at $z=140$ km in height. On the right are the standard binned and unfiltered $v_{z}$ power distributions. The black curves are drawn by eye to denote where the phase shifts would be in the power plots. Once again, the dashed vertical line is the position of the $a=c$ for the observational height.}\label{theory3}
\end{figure*}

In order to prove that the halo is produced by the return of reflected fast waves, we examine several intriguing features present in our simulations. Firstly, in a similar manner to \cite{rijs2015} we perform several identical simulations to the 2.7 kG case examined above, except with incrementally smaller Alfv\'en limiter values. \\
After undergoing mode conversion at around the $a=c$ height, fast magnetic waves will begin to refract and then ultimately reflect at the point in the atmosphere where the horizontal phase speed equals the Alfv\'en speed (i.e. where $\omega/k_{h}=a$).\\
By reducing the height of the artificial `cap' on the atmosphere we are allowing less and less room for fast waves to refract and deposit extra energy onto observable heights. Waves that impinge on the altered region of constant $a$ will simply travel upwards and out of the local area. As the original simulation had a value of $a_{lim}=90$ km/s, we run simulations with $a_{lim}=40$, 20 and 12 km/s and analyse the power in a similar manner to Figure \ref{obso2}, i.e. as a function of $|\mathbf{B}|$ and $\nu$. The comparison is shown in Figure \ref{theory1}.
\\
Panel d) corresponds to the case where the limiter is only barely above the $a=c$ height, enabling the mode conversion to take effect but yielding virtually no room for fast waves to return. Moreover in the intermediate cases of panels b) and c), the magnitude is reduced as the more vertically oriented waves are escaping to the top of the box, yielding contributions from only the more horizontally inclined waves.\\
Clearly the halo is entirely dependent on the overlying atmosphere and by restricting the refraction and return of the fast waves the enhancement is entirely absent. \\
The second theoretical check we perform is to compare the structure of the halo resulting from both the horizontal and vertical components of the velocity. A reasonable attack angle between the horizontal component of the wavevector, $k_{h}$ and $\mathbf{B}$ is still entirely likely in our simulations, as the field is never entirely horizontal. Also, as noted by \citet{khomenko2009} we can expect that it would at least be of a similar strength to the $v_{z}$ halo, as waves are largely horizontal at around the refraction height. 
\begin{figure*}
   \centering
    \includegraphics[width=1.0\textwidth,trim=0cm 1cm 0cm 1cm]{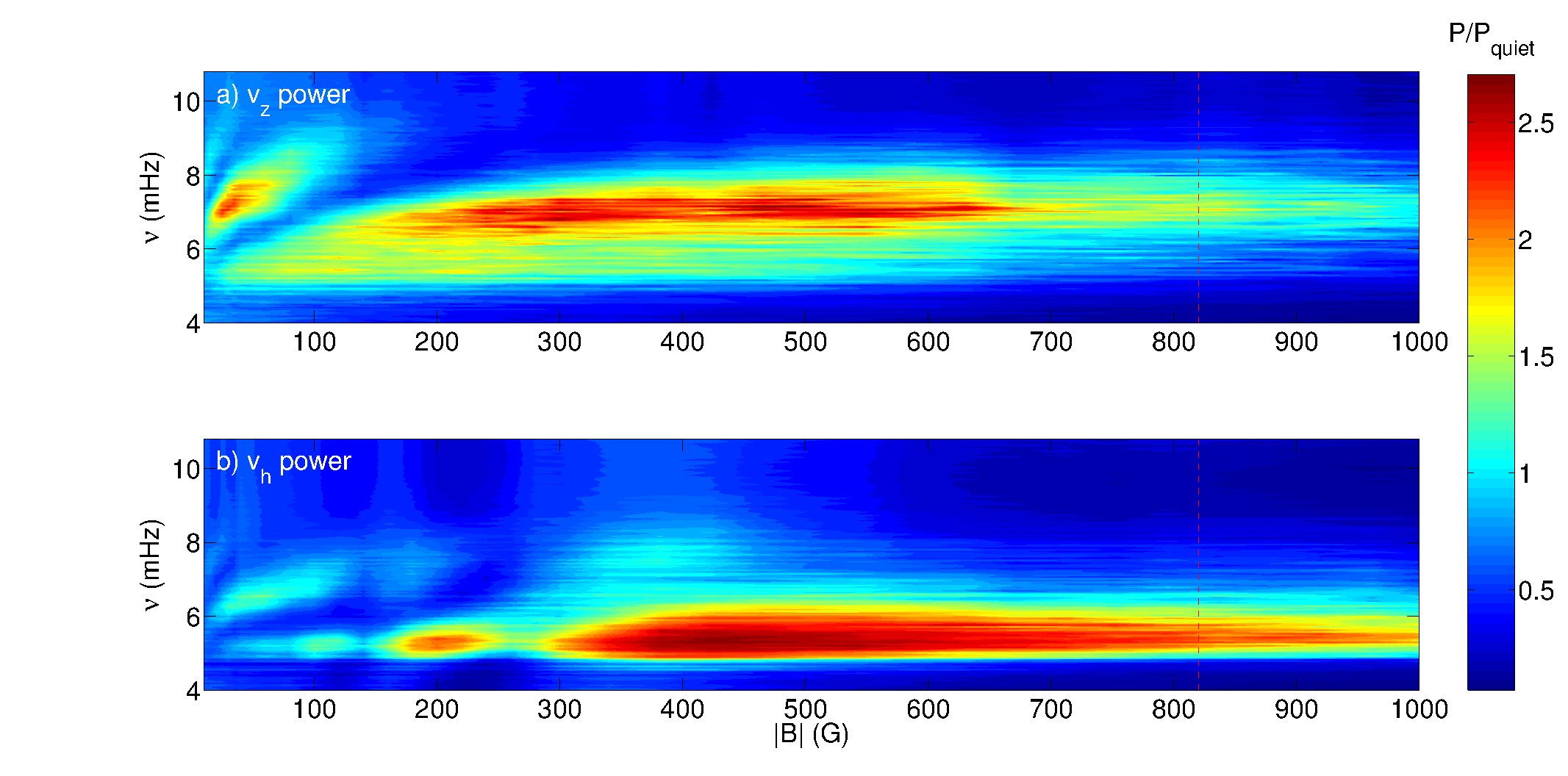}
    \caption{Panel a) is the standard binned $v_{z}$ power for the 2.7 kG atmosphere. Panel b) is the binned power corresponding to the horizontal component of velocity, $v_{h}$. The dashed vertical line is the position of the $a=c$ for the observational height of $z=140$ km. }\label{theory2}
\end{figure*}
\\
Figure \ref{theory2} shows the comparison. A clear feature is that the $v_{h}$ enhancement occurs at preferentially higher field strength than the $v_{z}$ enhancement. This feature also makes sense as the field inclination is more vertical at these radii, providing a larger attack angle.\\
It would be extremely useful if there were any center-to-limb observational studies of halo features, so that we could compare the horizontal Doppler component with our $v_{h}$. \citet{zharkov2013} have performed an analysis of the umbral "belly button" as a function of observation angle, but as of yet, no such studies focusing on halo properties have been conducted.\\
Finally, and perhaps most importantly, we explain the "dual-ring" power enhancement structure seen in the power maps earlier and in observations. \\
In Figure \ref{theory3} we compare $v_{z}$ power (once again at the standard observational height of 140 km) for both the 1.4 kG and the 2.7 kG simulations with the phase shifts at the same height. The phase shifts in this case are those calculated at $z=140$ km height, with respect to waves at $z=0$, so we are only looking at the phase shifts that the waves experience over a height change of 140 km in the simulation.
\\
The black curves have been added simply by eye to aid in the comparisons here. In both simulations there is a similar phase shift pattern to that observed in both the simulated and observed intensities at greater heights, however the magnitude is less here as the waves have travelled a shorter vertical distance. \\
The key fact to note is that the strong branch of positive phase shifts corresponds precisely to the region between the dual rings of power enhancement. This enhancement gap in $|\mathbf{B}|$, $\nu$ space is the dark `moat' seen between the two halo rings at various frequencies in the simulation power maps of Figure \ref{obso1}.\\
The halo itself shows no real phase shift which most likely indicates a mixture of upwards and downwards travelling waves. This is to be expected at high, non-trapped frequencies as waves rise upwards towards the $a=c$ layer and are refracted back downwards. The halo structure itself does not appear to change too significantly with respect to the peak magnetic field strength of the model, apart from the noted correlation with the $a=c$ layer. We certainly do not see any noticeable change in the peak halo frequency, as \citet{khomenko2009} suggested may be the case. This is most likely due to the fact that, although the peak field strengths of the two models are considerably different in the umbra (1.4 kG and 2.7 kG), at the halo radius (some 20 Mm out) the difference in the field strength will not be so significant.\\
The pertinent question is: why are there only upwards travelling waves in the moat in between the concentric halos?

\section{Fast wave damping and Alfv\'en waves}
The answer would appear to lie in the process of fast-Alfv\'en mode conversion, the basics of which are described in \citet{cally2011} and \citet{cally&hansen2011}. \\
Fast-Alfv\'en mode conversion has been well studied in both sunspot-like \citep{moradi2014,moradi2015} and simple magnetic field geometries: \citet{pascoe2011,pascoe2012} have studied the damping of transverse kink waves in terms of the associated Alfv\'en resonance and \citet{cally2008} and later \citet{khomenko&cally2011} have conducted parameter studies with monchromatic wave sources and simple inclined field magnetic structures. The finding of the latter two works was that fast wave energy is converted to the field-aligned Alfv\'en wave at favoured field inclinations ($\theta$) and wavevector-to-field angles ($\phi$). The process is also strongly dependent on both $\nu$ and $k_{h}$.\\
In the case of our distributed source simulations, waves exhibit a distribution of wavenumbers and frequencies in a similar manner to the quiet sun and so the picture is somewhat muddied in comparison to such simulations. We can expect however that fast-Alfv\'en conversion will in some way act on fast waves as they reach the Alfv\'en resonance near their upper turning point (on the order of a few hundred kilometres above the $a=c$, depending on $k_{h}$).  \\
As the halo appears to be generated by downwards turning fast waves, we would anticipate that some of this returning energy may be lost to the field aligned Alfv\'en wave, which will follow the local field lines until reaching the top (or the side) of the simulation domain.\\
In Figure \ref{obso1} we noted the strong concentric halos and the gap of power enhancement in between them. Figure \ref{theory3} shows this more comprehensively and associates this dark ring with a strong positive phase shift. \\
We suggest that the reason that the halo is not one continuous region is that for specific field inclinations, fast mode energy is lost to the Alfv\'en wave.
\begin{figure*}
   \centering
    \includegraphics[width=1.0\textwidth,trim=0cm 1cm 0cm 1cm]{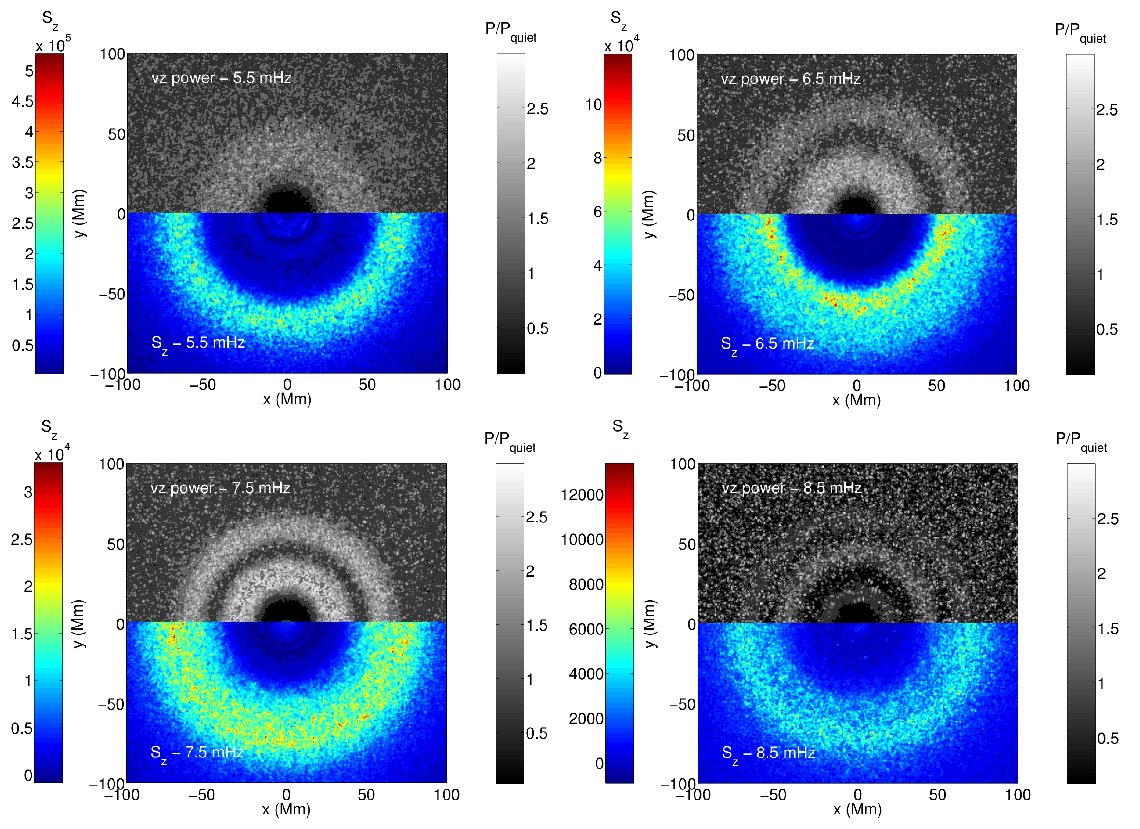}
\vskip 0.5 cm
    \caption{Power map - Poynting vector composites. Top halves are $xy$ power maps at $z=140$ km, filtered around the respective frequencies. Bottom halves are $S_{z}$ in units of ergs/$m^{2}s$, calculated at $z=2$ Mm. Note that the Poynting vector scaling is not consistent from plot to plot, as there is significalty less energy arriving at the top of the box for each subsequently higher non-trapped frequency range.}\label{theory4}
\end{figure*}
\\
Figure \ref{theory4} suggests this to be the case. Each panel of the figure corresponds to a specific frequency filtering. The top halves of the panels are the same as the panels in the middle row of Figure \ref{obso1}, i.e. filtered power maps corresponding to the stronger field 2.7 kG simulation at the Doppler velocity observational height of 140 km.\\
The bottom halves of the panels show the magnetic energy associated with the Alfv\'en wave in the form of the Poynting vector, $\mathbf{S}$, where  
\begin{equation}
\mathbf{S}=\frac{1}{\mu_{0}} (-\mathbf{v} \times \mathbf{B}) \times \mathbf{b},
\end{equation}
where $\mathbf{v}$ and $\mathbf{b}$ indicate the perturbations to the velocity and the background field respectively.
The bottom panels show the vertical component of the vector, $S_{z}$, corresponding to the upcoming Alfv\'en flux, and are calculated at the very top of the simulation domain, at a height $z=2$ Mm, just before the PML comes into effect at the top of the box. In each case the velocity has been pre-filtered around the associated frequency range prior to the calculation of $S_{z}$ to match the power maps.\\
It is worth remembering that we have applied a cap to $a$ above $a=90$ km/s in the atmosphere and so any upwards travelling Alfv\'en waves  will encounter our modified atmosphere and travel at a constant speed to the top of the box, instead of being subject to a rapidly increasing Alfv\'en speed. \\
The correlation between the Alfv\'en flux and the position of the dark ring is immediately noticeable, especially in the 5.5 and 6.5 mHz cases. Note that upwards travelling Alfv\'en waves will follow the field and that there is some field spreading with height in this MHS atmosphere which is why $S_{z}$ is diffuse and does not align precisely with the dark ring at observation heights.\\
It seems clear that this Alfv\'en wave energy is responsible for the strong band of positive phase shifts (and thus upwards travelling waves) in the dark moat. There is no wave energy left to return downwards at these radii and field inclinations. \\
Furthermore this supports the fast-wave halo mechanism rather strongly as the two processes are critically interlinked.\\

\section{Discussion and conclusion}

Linear forward modelling in realistic MHS sunspot atmospheres has yielded acoustic halos that match up quite well with observations, both spatially and spectrally. Apart from the magnitudes of the enhancements themselves, most observed features seem to be reproduced in our simulations, not just when comparing Doppler and vertical velocities, but also intensity halos at multiple heights in the chromosphere. As in the observations we see no power enhancement in calculations of the time-averaged intensity continuum power. \\
We have also presented convincing evidence that the mechanism responsible for halo formation is the refraction and return of magneto-acoustic fast waves at non-trapped frequencies. The halo appears very sensitive to the position of the $a=c$ layer in the atmosphere, which is the critical location for fast-wave mode conversion. With our realistic distributed wave source, we see a strong relationship between the strength of the halo and the extent to which fast waves are allowed to return downwards. This suggests that the halo is completely governed by the overlying $a > c$ atmosphere and the extra energy injected to observable heights by these returning fast waves.\\
The theory also predicts that an enhancement should be present in the power of $v_{h}$, as this component will also interact with the field, and that this enhancement should be concentrated toward more vertical field (as the horizontal component makes a larger attack angle with vertical field); this was shown to be the case as well. Unfortunately center-to-limb observational studies of the halo do not yet exist and so we cannot compare this horizontal velocity halo to the real thing. \\
Our simulations are performed in a MHS atmosphere, solving the linear MHD equations and using a wave excitation mechanism that approximates the wave bath of the solar photosphere. The fact that we see halos in such simulations (which are of course, entirely non-radiative and do not in any way include convective effects) suggests that the halo is not created by any convective cell-magnetic field interaction as suggested by \citet{jacoutot2008}.\\
Similarly, the idea of \citet{kuridze2008} that $m>1$ waves may become trapped in magnetic canopy structures cannot occur in our simulations as the field configuration is horizontally enforced at the boundaries and there is therefore no downwards oriented canopy.\\
The scattering mechanism of \cite{hanasoge2009} also cannot explain why the magnitude of the halo is determined entirely by the structure of the overlying atmosphere, as we have seen here.\\
As noted previously (and as can be seen in Figure \ref{obso2} in particular), the primary difference between our simulated halos and those actually observed in the photosphere and chromosphere is the magnitude of the enhancement itself. Observed Doppler velocity halos have magnitudes up to 60\% (over the quiet sun average at the same height). Our simulated $v_{z}$ halos are greater than this by a factor of 2 or even 3, depending on frequency. \\
There are several possible explanations for this discrepancy. As we have postulated, the halo enhancement most likely occurs as a result of fast waves interacting with the sunspot magnetic field at large attack angles. This yields a large conversion fraction to the fast magnetic wave which refracts and deposits additional energy in the photopshere and chromosphere. The penumbral field structure of active regions differs significantly from the simple MHS model used here however. Our atmosphere does not explicitly include an umbra or penumbra, but rather consists of a smoothly decreasing field strength and vertical inclination component, yielding significant regions of smooth, nearly-horizontal field. Non-trapped waves which reach the $a=c$ equipartition layer will have a large vertical component and so we would expect these waves to interact strongly with primarily horizontal field. In nature, penumbrae contain fine structure, with bright and dark filaments giving rise to the now well-observed combed magnetic field configuration \citep{scharmer2002,bellotrubio2004}. At the outer penumbral boundary, studies have shown up to a 60$^{\circ}$ difference in field inclination between dark (largely horizontal) and light (largely vertical) filaments \citep{weiss2004,thomas2006}. Energy corresponding to waves interacting with nearly vertical field at these radii would therefore be lost, transmitting primarily to the slow magneto-acoustic mode. This would have an overall effect of weakening the halo, and as these features are not represented in our model, they may be a contributing factor for our high halo magnitudes.\\
Another factor to consider is the non-ideal nature of the photosphere, which contains a large neutral component \citep{krasno2010,khomenko2012}. In our ideal MHD assumption we assume full ionization, and thus neglect any dissipative effects brought about by ion-neutral collisions. It is conceivable that these partial ionization dissipative effects (as well as any other dissipation brought about by small scale magnetic structure) in the real photosphere and chromosphere may reduce the observed velocity and intensity halos.\\
With regards to intensity halos, Figure \ref{obso2} shows a good agreement between the magnitudes of the observed and simulated AIA 1700 {\AA} halo. However the observed AIA 1600 {\AA} halo is very weak, in contrast to the simulation. This may be due to the larger height range over which observational intensities are calculated. In particular, the height over which the AIA 1600 {\AA} intensity band is determined observationally is some 185 km (centered at 430 km in height) \citep{fossum2005}, which may have the effect of smoothing out the 1600 {\AA} intensity power, given that the corresponding synthetic intensities encompass a much narrower height range.\\
In our final discovery of note we have shown that not just fast-slow mode conversion but also fast-Alfv\'en conversion plays a considerable role in the formation of the halo. This conversion of the fast wave at preferential field inclinations takes energy away along the field lines in the form of the transverse Alfv\'en wave, resulting in the dual-ring halo structure seen at high frequency. In our simulations this is visible at 6.5 mHz and above - the halo is essentially being broken up into two concentric rings by this Alfv\'enic energy loss. Observationally this may help to explain the underlying process responsible for the 8-9 mHz dual-ring power halo structure \citep{hanson2015,rajaguru2012}, with its spatially localized zone of enhancement, dark moat and diffuse enhancement region structure.

\begin{acknowledgments}
The above work would not have been possible without the generous computing time provided by the center for Astrophysics and Supercomputing at Swinburne University of Technology (Australia), the Multi-modal Australian ScienceS Imaging and Visualisation Environment (MASSIVE; www.massive.org.au) and the NCI National Facility systems at the Australian National University.
\end{acknowledgments}

\bibliographystyle{apj}
\bibliography{database}

\begin{thebibliography}{58}
\expandafter\ifx\csname natexlab\endcsname\relax\def\natexlab#1{#1}\fi

\bibitem[{{Avrett}(1981)}]{avrett1981}
{Avrett}, E.~H. 1981, in The Physics of Sunspots, ed. L.~E. {Cram} \& J.~H.
  {Thomas}, 235--255

\bibitem[{{Bellot Rubio} {et~al.}(2004){Bellot Rubio}, {Balthasar}, \&
  {Collados}}]{bellotrubio2004}
{Bellot Rubio}, L.~R., {Balthasar}, H., \& {Collados}, M. 2004, A\&A, 427, 319

\bibitem[{{Bogdan} {et~al.}(1996){Bogdan}, {Hindman}, {Cally}, \&
  {Charbonneau}}]{bogdan1996}
{Bogdan}, T.~J., {Hindman}, B.~W., {Cally}, P.~S., \& {Charbonneau}, P. 1996,
  ApJ, 465, 406

\bibitem[{{Braun} {et~al.}(1987){Braun}, {Duvall}, \& {Labonte}}]{braun1987}
{Braun}, D.~C., {Duvall}, Jr., T.~L., \& {Labonte}, B.~J. 1987, ApJL, 319, L27

\bibitem[{{Braun} {et~al.}(1992){Braun}, {Lindsey}, {Fan}, \&
  {Jefferies}}]{braun1992}
{Braun}, D.~C., {Lindsey}, C., {Fan}, Y., \& {Jefferies}, S.~M. 1992, ApJ, 392,
  739

\bibitem[{{Brown} {et~al.}(1992){Brown}, {Bogdan}, {Lites}, \&
  {Thomas}}]{brown1992}
{Brown}, T.~M., {Bogdan}, T.~J., {Lites}, B.~W., \& {Thomas}, J.~H. 1992, ApJL,
  394, L65

\bibitem[{{Cally}(2006)}]{cally2006}
{Cally}, P.~S. 2006, Royal Society of London Philosophical Transactions Series
  A, 364, 333

\bibitem[{{Cally}(2007)}]{cally2007}
---. 2007, Astronomische Nachrichten, 328, 286

\bibitem[{{Cally}(2011)}]{cally2011}
{Cally}, P.~S. 2011, in Astronomical Society of India Conference Series,
  Vol.~2, Astronomical Society of India Conference Series, 221--227

\bibitem[{{Cally} \& {Bogdan}(1997)}]{cally1997}
{Cally}, P.~S., \& {Bogdan}, T.~J. 1997, ApJL, 486, L67

\bibitem[{{Cally} {et~al.}(2003){Cally}, {Crouch}, \& {Braun}}]{cally2003}
{Cally}, P.~S., {Crouch}, A.~D., \& {Braun}, D.~C. 2003, MNRAS, 346, 381

\bibitem[{{Cally} \& {Goossens}(2008)}]{cally2008}
{Cally}, P.~S., \& {Goossens}, M. 2008, SoPh, 251, 251

\bibitem[{{Cally} \& {Hansen}(2011)}]{cally&hansen2011}
{Cally}, P.~S., \& {Hansen}, S.~C. 2011, ApJ, 738, 119

\bibitem[{{Cally} \& {Moradi}(2013)}]{cally2013}
{Cally}, P.~S., \& {Moradi}, H. 2013, MNRAS, 435, 2589

\bibitem[{{Christensen-Dalsgaard} {et~al.}(1996){Christensen-Dalsgaard},
  {Dappen}, {Ajukov}, {Anderson}, {Antia}, {Basu}, {Baturin}, {Berthomieu},
  {Chaboyer}, {Chitre}, {Cox}, {Demarque}, {Donatowicz}, {Dziembowski},
  {Gabriel}, {Gough}, {Guenther}, {Guzik}, {Harvey}, {Hill}, {Houdek},
  {Iglesias}, {Kosovichev}, {Leibacher}, {Morel}, {Proffitt}, {Provost},
  {Reiter}, {Rhodes}, {Rogers}, {Roxburgh}, {Thompson}, \&
  {Ulrich}}]{dalsgaard1996}
{Christensen-Dalsgaard}, J., {et~al.} 1996, Science, 272, 1286

\bibitem[{{Finsterle} {et~al.}(2004){Finsterle}, {Jefferies}, {Cacciani},
  {Rapex}, \& {McIntosh}}]{finsterle2004}
{Finsterle}, W., {Jefferies}, S.~M., {Cacciani}, A., {Rapex}, P., \&
  {McIntosh}, S.~W. 2004, ApJL, 613, L185

\bibitem[{{Fleck} {et~al.}(2011){Fleck}, {Couvidat}, \& {Straus}}]{fleck2011}
{Fleck}, B., {Couvidat}, S., \& {Straus}, T. 2011, SoPh, 271, 27

\bibitem[{{Fossum} \& {Carlsson}(2005)}]{fossum2005}
{Fossum}, A., \& {Carlsson}, M. 2005, ApJ, 625, 556

\bibitem[{{Hanasoge}(2007)}]{hanasogephd2007}
{Hanasoge}, S.~M. 2007, PhD thesis, Stanford University

\bibitem[{{Hanasoge}(2009)}]{hanasoge2009}
---. 2009, A\&A, 503, 595

\bibitem[{{Hanasoge} {et~al.}(2007){Hanasoge}, {Duvall}, \&
  {Couvidat}}]{hanasoge2007}
{Hanasoge}, S.~M., {Duvall}, Jr., T.~L., \& {Couvidat}, S. 2007, ApJ, 664, 1234

\bibitem[{{Hanson} {et~al.}(2015){Hanson}, {Donea}, \& {Leka}}]{hanson2015}
{Hanson}, C.~S., {Donea}, A.~C., \& {Leka}, K.~D. 2015, SoPh

\bibitem[{{Hindman} \& {Brown}(1998)}]{hindman1998}
{Hindman}, B.~W., \& {Brown}, T.~M. 1998, ApJ, 504, 1029

\bibitem[{{Jacoutot} {et~al.}(2008){Jacoutot}, {Kosovichev}, {Wray}, \&
  {Mansour}}]{jacoutot2008}
{Jacoutot}, L., {Kosovichev}, A.~G., {Wray}, A., \& {Mansour}, N.~N. 2008,
  ApJL, 684, L51

\bibitem[{{Jain} \& {Haber}(2002)}]{jain2002}
{Jain}, R., \& {Haber}, D. 2002, A\&A, 387, 1092

\bibitem[{{Jess} {et~al.}(2012){Jess}, {Shelyag}, {Mathioudakis}, {Keys},
  {Christian}, \& {Keenan}}]{jess2012}
{Jess}, D.~B., {Shelyag}, S., {Mathioudakis}, M., {Keys}, P.~H., {Christian},
  D.~J., \& {Keenan}, F.~P. 2012, ApJ, 746, 183

\bibitem[{{Khomenko} \& {Cally}(2011)}]{khomenko&cally2011}
{Khomenko}, E., \& {Cally}, P.~S. 2011, Journal of Physics Conference Series,
  271, 012042

\bibitem[{{Khomenko} \& {Collados}(2008)}]{khomenko2008}
{Khomenko}, E., \& {Collados}, M. 2008, ApJ, 689, 1379

\bibitem[{{Khomenko} \& {Collados}(2009)}]{khomenko2009}
---. 2009, A\&A, 506, L5

\bibitem[{{Khomenko} \& {Collados}(2012)}]{khomenko2012}
---. 2012, ApJ, 747, 87

\bibitem[{{Krasnoselskikh} {et~al.}(2010){Krasnoselskikh}, {Vekstein},
  {Hudson}, {Bale}, \& {Abbett}}]{krasno2010}
{Krasnoselskikh}, V., {Vekstein}, G., {Hudson}, H.~S., {Bale}, S.~D., \&
  {Abbett}, W.~P. 2010, ApJ, 724, 1542

\bibitem[{{Kuridze} {et~al.}(2008){Kuridze}, {Zaqarashvili}, {Shergelashvili},
  \& {Poedts}}]{kuridze2008}
{Kuridze}, D., {Zaqarashvili}, T.~V., {Shergelashvili}, B.~M., \& {Poedts}, S.
  2008, Annales Geophysicae, 26, 2983

\bibitem[{{Kurucz}(1993)}]{kurucz1993}
{Kurucz}, R. 1993, ATLAS9 Stellar Atmosphere Programs and 2 km/s grid.~Kurucz
  CD-ROM No.~13.~ Cambridge, Mass.: Smithsonian Astrophysical Observatory,
  1993., 13

\bibitem[{{Lemen} {et~al.}(2012){Lemen}, {Title}, {Akin}, {Boerner}, {Chou},
  {Drake}, {Duncan}, {Edwards}, {Friedlaender}, {Heyman}, {Hurlburt}, {Katz},
  {Kushner}, {Levay}, {Lindgren}, {Mathur}, {McFeaters}, {Mitchell}, {Rehse},
  {Schrijver}, {Springer}, {Stern}, {Tarbell}, {Wuelser}, {Wolfson}, {Yanari},
  {Bookbinder}, {Cheimets}, {Caldwell}, {Deluca}, {Gates}, {Golub}, {Park},
  {Podgorski}, {Bush}, {Scherrer}, {Gummin}, {Smith}, {Auker}, {Jerram},
  {Pool}, {Soufli}, {Windt}, {Beardsley}, {Clapp}, {Lang}, \&
  {Waltham}}]{lemen2012}
{Lemen}, J.~R., {et~al.} 2012, SoPh, 275, 17

\bibitem[{{Low}(1980)}]{low1980}
{Low}, B.~C. 1980, SoPh, 67, 57

\bibitem[{{Moradi} \& {Cally}(2013)}]{moradi2013}
{Moradi}, H., \& {Cally}, P.~S. 2013, Journal of Physics Conference Series,
  440, 012047

\bibitem[{{Moradi} \& {Cally}(2014)}]{moradi2014}
---. 2014, ApJL, 782, L26

\bibitem[{{Moradi} {et~al.}(2015){Moradi}, {Cally}, {Przybylski}, \&
  {Shelyag}}]{moradi2015}
{Moradi}, H., {Cally}, P.~S., {Przybylski}, D., \& {Shelyag}, S. 2015, MNRAS,
  449, 3074

\bibitem[{{Moretti} {et~al.}(2007){Moretti}, {Jefferies}, {Armstrong}, \&
  {McIntosh}}]{moretti2007}
{Moretti}, P.~F., {Jefferies}, S.~M., {Armstrong}, J.~D., \& {McIntosh}, S.~W.
  2007, A\&A, 471, 961

\bibitem[{{Parchevsky} \& {Kosovichev}(2007)}]{parchevsky2007}
{Parchevsky}, K.~V., \& {Kosovichev}, A.~G. 2007, ApJ, 666, 547

\bibitem[{{Pascoe} {et~al.}(2012){Pascoe}, {Hood}, {de Moortel}, \&
  {Wright}}]{pascoe2012}
{Pascoe}, D.~J., {Hood}, A.~W., {de Moortel}, I., \& {Wright}, A.~N. 2012,
  A\&A, 539, A37

\bibitem[{{Pascoe} {et~al.}(2011){Pascoe}, {Wright}, \& {De
  Moortel}}]{pascoe2011}
{Pascoe}, D.~J., {Wright}, A.~N., \& {De Moortel}, I. 2011, ApJ, 731, 73

\bibitem[{{Pizzo}(1986)}]{pizzo1986}
{Pizzo}, V.~J. 1986, ApJ, 302, 785

\bibitem[{{Przybylski} {et~al.}(2015){Przybylski}, {Shelyag}, \&
  {Cally}}]{przybylski2015}
{Przybylski}, D., {Shelyag}, S., \& {Cally}, P.~S. 2015, ApJ, 807, 20

\bibitem[{{Rajaguru} {et~al.}(2013){Rajaguru}, {Couvidat}, {Sun}, {Hayashi}, \&
  {Schunker}}]{rajaguru2012}
{Rajaguru}, S.~P., {Couvidat}, S., {Sun}, X., {Hayashi}, K., \& {Schunker}, H.
  2013, SoPh, 287, 107

\bibitem[{{Rempel} {et~al.}(2009){Rempel}, {Sch{\"u}ssler}, \&
  {Kn{\"o}lker}}]{rempel2009}
{Rempel}, M., {Sch{\"u}ssler}, M., \& {Kn{\"o}lker}, M. 2009, ApJ, 691, 640

\bibitem[{{Rijs} {et~al.}(2015){Rijs}, {Moradi}, {Przybylski}, \&
  {Cally}}]{rijs2015}
{Rijs}, C., {Moradi}, H., {Przybylski}, D., \& {Cally}, P.~S. 2015, ApJ, 801,
  27

\bibitem[{{Scharmer} {et~al.}(2002){Scharmer}, {Gudiksen}, {Kiselman},
  {L{\"o}fdahl}, \& {Rouppe van der Voort}}]{scharmer2002}
{Scharmer}, G.~B., {Gudiksen}, B.~V., {Kiselman}, D., {L{\"o}fdahl}, M.~G., \&
  {Rouppe van der Voort}, L.~H.~M. 2002, Nat, 420, 151

\bibitem[{{Scherrer} {et~al.}(1995){Scherrer}, {Bogart}, {Bush}, {Hoeksema},
  {Kosovichev}, {Schou}, {Rosenberg}, {Springer}, {Tarbell}, {Title},
  {Wolfson}, {Zayer}, \& {MDI Engineering Team}}]{scherrer1995}
{Scherrer}, P.~H., {et~al.} 1995, SoPh, 162, 129

\bibitem[{{Scherrer} {et~al.}(2012){Scherrer}, {Schou}, {Bush}, {Kosovichev},
  {Bogart}, {Hoeksema}, {Liu}, {Duvall}, {Zhao}, {Title}, {Schrijver},
  {Tarbell}, \& {Tomczyk}}]{scherrer2012}
---. 2012, SoPh, 275, 207

\bibitem[{{Schunker} \& {Braun}(2011)}]{schunker2011}
{Schunker}, H., \& {Braun}, D.~C. 2011, SoPh, 268, 349

\bibitem[{{Schunker} \& {Cally}(2006)}]{schunker2006}
{Schunker}, H., \& {Cally}, P.~S. 2006, MNRAS, 372, 551

\bibitem[{{Spruit} \& {Bogdan}(1992)}]{spruit1992}
{Spruit}, H.~C., \& {Bogdan}, T.~J. 1992, ApJL, 391, L109

\bibitem[{{Thomas} {et~al.}(2006){Thomas}, {Weiss}, {Tobias}, \&
  {Brummell}}]{thomas2006}
{Thomas}, J.~H., {Weiss}, N.~O., {Tobias}, S.~M., \& {Brummell}, N.~H. 2006,
  A\&A, 452, 1089

\bibitem[{{Toner} \& {Labonte}(1993)}]{toner1993}
{Toner}, C.~G., \& {Labonte}, B.~J. 1993, ApJ, 415, 847

\bibitem[{{Vernazza} {et~al.}(1981){Vernazza}, {Avrett}, \&
  {Loeser}}]{vernazza1981}
{Vernazza}, J.~E., {Avrett}, E.~H., \& {Loeser}, R. 1981, ApJS, 45, 635

\bibitem[{{Weiss} {et~al.}(2004){Weiss}, {Thomas}, {Brummell}, \&
  {Tobias}}]{weiss2004}
{Weiss}, N.~O., {Thomas}, J.~H., {Brummell}, N.~H., \& {Tobias}, S.~M. 2004,
  ApJ, 600, 1073

\bibitem[{{Zharkov} {et~al.}(2009){Zharkov}, {Shelyag}, {Fedun}, {Erd{\'e}lyi},
  \& {Thompson}}]{zharkov2013}
{Zharkov}, S., {Shelyag}, S., {Fedun}, V., {Erd{\'e}lyi}, R., \& {Thompson},
  M.~J. 2009, ArXiv e-prints

\end{thebibliography}

\end{document}